\documentclass[useAMS,usenatbib,usegraphicx]{mn2e}

\def\ha{H$\alpha$ $\lambda 6563$}
\def\hb{H$\beta$ $\lambda 4861$}
\def\nii{[NII] $\lambda 6583$}

\def\kms{$\mathrm {km\,s}^{-1}$}
\def\mdot{$\mathrm {M_{\odot}\,yr}^{-1}$}
\def\msun{$\mathrm {M_{\odot}}$}
\def\density{$10^{-6}\ \mathrm {M_{\odot}\,pc}^{-3}$}
\def\oversim#1#2{\lower0.5pt\vbox{\baselineskip0pt \lineskip-0.5pt
     \ialign{$\mathsurround0pt #1\hfil##\hfil$\crcr#2\crcr\sim\crcr}}}
\def\lsim{\mathrel{\mathpalette\oversim<}}    

\title[The shaping of PNe through interaction with the ISM]{The shaping of planetary nebula Sh\,2-188 through interaction with the interstellar medium}
\author[C. J. Wareing et al.]{C. J. Wareing$^{1}$\thanks{E-mail: cwareing@jb.man.ac.uk}, 
T. J. O'Brien$^{1}$, 
Albert A. Zijlstra$^{2}$, 
K. B. Kwitter$^{3}$,
J. Irwin$^4$,
\newauthor
N. Wright$^5$,
R. Greimel$^6$ and 
J. E. Drew$^{7}$\\
$^{1}$Jodrell Bank Observatory, University of Manchester, Macclesfield, Cheshire, SK11 9DL, UK\\
$^{2}$School of Physics and Astronomy, University of Manchester, Oxford Road, Manchester, M13 9PL, UK\\
$^{3}$Department of Astronomy, Williams College, 33 Lab Campus Drive, Williamstown, MA 01267, USA\\
$^{4}$Institute of Astronomy, Cambridge University, Madingley Road, Cambridge, CB3 0HA, UK\\
$^{5}$University College London, Department of Physics \& Astronomy, Gower Street, London, WC1E 6BT, UK\\
$^{6}$Isaac Newton Group of Telescopes, Apartado de corras 321, E-38700 Santa Cruz de La Palma, Tenerife, Spain\\
$^{7}$Imperial College of Science, Technology and Medicine, Blackett Laboratory, Prince Consort Road, London, SW7 2BW, UK}
\begin{document}

\date{in original form 2005 Sep 6}


\maketitle

\label{firstpage}

\begin{abstract}
  
  Sh\,2-188 is an example of strong interaction between a planetary nebula (PN) and
  the interstellar medium (ISM). It shows a single arc-like structure, consisting of several
  filaments, which is postulated to be the result of motion through the ISM.
  We present new H$\alpha$ images from the Isaac Newton Telescope Photometric H$\alpha$  
  Survey of the Northern Galactic Plane (IPHAS) which reveal structure behind the
  filamentary limb. A faint, thin arc is seen opposite the bright limb, in
  combination forming a closed ring.  Behind the faint arc a long wide tail
  is detected, doubling the size of the nebula. The nebula extends 15 arcmin
  on the sky in total. We have developed a `triple-wind' hydrodynamical
  model, comprising of the initial `slow' asymptotic giant branch (AGB) wind and the later `fast'
  stellar wind (the interacting stellar wind model), plus a third wind
  reflecting the motion through the ISM. Simulations at various velocities of
  the central star relative to the ISM indicate that a high velocity of 125 \kms\ 
  is required to reproduce the observed structure. We find that the bright
  limb and the tail already formed during the AGB phase, prior to the formation
  of the PN. The closure of the ring arises from the slow--fast wind
  interaction. Most of the mass lost on the AGB has been swept downstream,
  providing a potential explanation of the missing mass problem in PNe. We report a
  proper motion for the central star of $30\pm10$ mas\,yr$^{-1}$ in the
  direction of the bright limb. Assuming the central star is moving at
  $125\pm25$ \kms, the distance to the nebula is estimated to be
  $850^{+500}_{-420}$ pc, consistent with a spectroscopic distance to the
  star. Expansion velocities measured from spectroscopic data of the bright filaments
  are consistent with velocities measured from the simulation. Sh\,2-188 is 
  one of the largest PNe known, with an extent of 2.8 
  pc. The model shows that this size was already set during the AGB phase.

\end{abstract}


\section{Introduction}

The accepted theory of planetary nebula (PN) formation is the interacting
stellar winds model (ISW) \citep{kwok82,balick87} where a fast wind ($\sim10^3$ \kms)
from the hot central star of a PN blows into the slow wind ($\sim10$ \kms)
produced during the preceding asymptotic giant branch (AGB) phase.  The inner regions of the
slow wind are compressed into a dense shell and ionised by the energetic UV
radiation of the central star. The familiar ring-like appearance of PNe is
then observed. Structures in the nebula are normally attributed to asymmetries
in the slow wind, related to physical properties of the central star, such as
rotation or binarity.  

Observations of PNe have shown several cases where the outer shell shows the
only departure from symmetry. For those cases, the cause of the asymmetries
has been proposed to be an interaction with the interstellar medium (ISM).
Interaction of PNe with the ISM was first discussed by \cite{gurzadyan69}. An
early theoretical study by \cite{smith76} assumed a thin shell approximation
and the `snowplough' model of \cite{oort51}. \cite{isaacmann79} used the
same approximation with higher velocities and ISM densities. Both of these
studies concluded similarly that a nebula fades away before any disruption of
the nebular shell becomes noticeable.

In contrast, \cite{borkowski90} found that many PNe with large angular extent
show signs of PN--ISM interaction, and that all nebulae containing central
stars with a proper motion greater than 0.015 arcsec\,yr$^{-1}$ do so.
\cite{soker91} hydrodynamically modelled the interaction. The PN shell is
first compressed in the direction of motion and then in later stages this part
of the shell is significantly decelerated with respect to the central star.
Both conclude that the interaction with the ISM becomes dominant when
the density of the nebular shell drops below a certain critical limit, of
typically $n_{\rm H} = 40 \rm cm^{-3}$ for a PN in the Galactic Plane.

\cite{villaver03} (hereafter referred to as VGM) pointed out the PN--ISM
interaction had only been studied by considering the relative movement when
the nebular shell had already formed.  They performed 2D hydrodynamic
simulations following the full AGB phase followed by the PN phase
\citep{vassiliadis93,vassiliadis94}, with a conservative relative velocity of
the central star of 20 \kms\ and conservative conditions of the surrounding
ISM of $n_{\rm H} =0.1 \rm cm^{-3}$.  VGM concluded that interaction provides
an adequate mechanism to explain the high rate of observed asymmetries in the
external shells of PNe. Further, they conclude that stripping of mass
downstream during the AGB phase provides a possible solution to the problem of
missing mass in PN whereby only a small fraction of the mass ejected during
the AGB phase is inferred to be present during the post-AGB phase.
Observational evidence for the effect of the ISM on AGB wind stuctures was
found by \cite{zijlstra02}.

The PN Sh\,2-188 \citep{sharpless59} is among the most extreme examples of ISM
interaction.  The nebula has a one-sided (semi-circular), filamentary
appearance.  It is a large nebula, with a reported 340 arcsec diameter
\citep{acker92}; the currrent paper shows it to be considerably larger. It is
located in the Galactic plane at $l = 128^\circ$, $b = -4^\circ$.  New data
are presented showing the faint back of the shell and an extended H$\alpha$
tail.  The unusual appearance suggests a high proper motion and makes Sh\,2-188
an important test case for PN--ISM interaction at high velocity.  We have
developed a `triple-wind' model using a initial slow AGB wind, a subsequent
fast post-AGB (PN) wind, and adding a third wind reflecting the movement
through the ISM into the ISW model. We use a hydrodynamic scheme developed by
\cite{wareing05}, to investigate whether this triple-wind model can reproduce
the nebular shape of Sh\,2-188 without requiring magnetic fields. We support
the results of the model with a proper motion study of the candidate central
star.

\section{Observations}

\subsection{Sh\,2-188}

Sh\,2-188 (PNG 128.0$-$04.1, also known as Simeiz 22) was photographed in 1965
by \cite{gaze65}.  Spectroscopic work
\citep{parker64,lozinskaya71,johnson75,kwitter79} revealed the object to have
an extremely high \nii\ / \ha\ line-ratio. Further spectroscopic data on the
filamentary structure \citep{rosado82} indicated filament densities of a few
hundred H cm$^{-3}$ and found abundances similar to Peimbert Type I PN, where
the large over-abundances of He and N result from CNO enrichment of the
stellar envelope prior to shell ejection \citep{peimbert81}. Mapping at radio
wavelengths \citep{israel76,salter84} found only a weak source which appeared
to have a flat (thermal) spectrum. This observational evidence, along with the
relatively bright optical appearance and PN abundances, constituted a
convincing argument to classify the object as a PN, rather than
a supernova remnant where non-thermal radio emission would be expected.

\cite{kwitter88} identified the candidate central star of the nebula with
geometric, colour and apparent magnitude methods. The star is displaced from
the geometric centre of the nebula. They estimate a percentage probability of
finding an unrelated blue star in the nebula to be 1-2 per cent at most.  In their
spectroscopic investigation of central stars of old PNe, \cite{napiwotzki95}
classified the candidate central star as hydrogen-rich.  \cite{borkowski90}
suggested that the filaments are most probably thin, sheetlike layers of
shock-compressed nebular gas, parallel to the line of sight. The majority of
the filaments are aligned along the crescent-shaped outer rim, but they noted
a few appear to be along the nebular symmetry axis, possibly indicating
stellar motion in that direction, as similar filaments do in Abell 35.

\begin{table}
\caption{Estimates of the distance to, radius of and age of Sh\,2-188.}
\label{distances}
\begin{center}
\begin{tabular}{|c|c|c|c|} \hline
$D$ & $R$ & $t$ & Source  \\
pc & pc & kyrs & \\
\hline
600 & 0.5 & 12.1 & \cite{napiwotzki95} \\
$800\pm300$ & & 26.0 & \cite{saurer95} \\
$965^{+1000}_{-600}$ & 0.88 & 22 & \cite{napiwotzki99,napiwotzki01} \\
\hline
\end{tabular}
\end{center}
\end{table}

Table \ref{distances} shows various recent estimates of the distance and physical parameters of the nebula. Various historical estimates can be found in table 5 of \cite{saurer95}. We do not list statistical distances \citep[e.g.][]{cahn92}: the Shklovski method (constant PN mass) particularly is of too low accuracy. \cite{saurer95} calculated an extinction distance based on published UBV photometry of stars in the angular vicinity used to construct a reddening distance relationship for the line of sight towards the nebula. Most recently, the NLTE modelling method of \cite{napiwotzki99,napiwotzki01} has implied the nebula is ancient with the central star at the turning point before the UV flux switches off and the star enters the white dwarf cooling regime. The parameters of Sh\,2-188 are still subject to large uncertainties. Consistently though, this nebula is found to be an evolved object and as such is included in the Atlas of Ancient PNe \citep{tweedy96}. 

\citet{tweedy96} note the existence of faint structure to the north-west of the bright nebula. They favour a simple explanation that the gas in this direction has diffused away. They infer a highly inhomogeneous ISM to the south-east producing the bright filaments but essentially absent to the north-west. A more likely model is one of a high peculiar velocity through a uniform ISM, and this is explored below.

\begin{figure}
\begin{center}
\includegraphics[width=7cm]{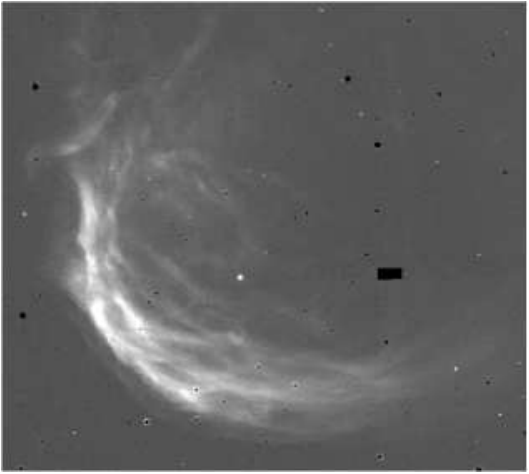}
\caption{An image of Sh\,2-188 created by a mosaic of images taken as part of the IPHAS survey in 2003 (see text). The faint central star is approximately 0.5 arcmin East of the brightest star in the centre of the nebula. North is up and East to the left. The image is $9.0 \times 8.5$ arcmin.}
\label{iphas-bow}
\end{center}
\end{figure}

\begin{figure}
\begin{center}
\includegraphics[width=7cm]{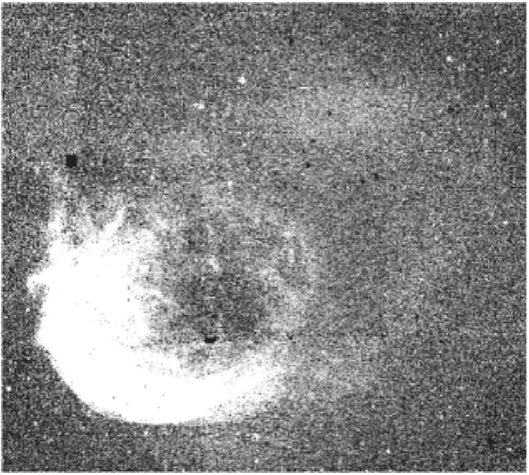}
\caption{Same as Fig. \ref{iphas-bow} but with scaling chosen to reveal faint structure. North is up and East to the left. The image shows a faint ring completing the bright semi-circular arc and tails stretching away connecting in the north-west. The ring is centred approximately at RA = $1^{\rm h}$ 30$^{\rm m}$ 25$^{\rm s}$, Dec = +58$^{\rm o}$\ 25$^\prime$ 30$^{\prime\prime}$ (J2000). The image is $18.0 \times 17.0$ arcmin.}
\label{iphas-deep}
\end{center}
\end{figure}

\subsection{H$\alpha$ observations}

Fig. \ref{iphas-bow} shows an image of Sh\,2-188 created by a mosaic of new
\ha\ observations taken as part of the Isaac Newton Telescope Photometric
H$\alpha$ Survey of the Northern Galactic Plane (IPHAS) \citep{drew05}. The IPHAS
survey entails wide-field imaging through H$\alpha$, r$^\prime$ and
i$^\prime$-band filters. It uses the Wide-Field Camera (WFC) at the 2.5-m INT,
located at La Palma. The goal of the survey is to image the entire northern
Galactic plane between $-5^{\rm o}$ and $+5^{\rm o}$, to study emission-line
stars and nebulae. The WFC consists of four 4k$\times$2k CCDs, in an L-shape
configuration.  The pixel scale is 0.33 arcsec and the instantaneous field is
about 0.3 square degrees.  Exposure times are 120 s for H$\alpha$ and 10 s
for i$^\prime$ and r$^\prime$ for the observations presented here. Each
pointing is supplemented by a second one offset by 5 arcmin SW, to cover the
gaps between the CCDs. The pointings together provide a tiled pattern that
minimizes the gaps, but because of the L-shape CCD configuration small gaps do
appear.

The area around Sh\,2-188 was observed on 2003 Oct 13 and 14. The seeing
varied between 1 and 1.5 arcsec. The airmass was between 1.2 and 1.4. Data
reduction was done using the pipeline procedures, which correct for the bias,
flat-fielding, non-linearity and geometric distortions. A detailed description
can be found in \citet{drew05}. Fields around the position of Sh\,2-188 were
retrieved and mosaiced together using the Virtual Observatory software package
{\sc MONTAGE}. The pixel scale was binned to 1 arcsec.  A 40$^\prime$ field
was mosaiced. The r$^\prime$ was subtracted from the H$\alpha$ image to bring
out the faint emission in this crowded field. {\sc MONTAGE} tends to leave
gradients across the full field, an artifact of the process of background
matching of fields taken with different sky brightness. This linear gradient
was subtracted.

The final image in Fig. \ref{iphas-bow} shows the bright portion of the nebula
in the south-east, which describes a semi-circular arc with a diameter of 10
arcmin. Due to seeing variations between the r$^\prime$ and H$\alpha$ frames,
small core-halo residuals can be seen at the positions of bright stars. Stars
can also leave positive or negative residuals due to the stellar absorption
features in the H$\alpha$ and/or r$^\prime$ filter. The bright residual star
just behind the brightest part of the shell is an M5 giant, where the
r$^\prime$ filter covers a TiO absorption band. We stress that this is not an
emission-line star, nor is it the central star of the PN. The central star of
the PN is approximately 0.5 arcmin East of this star. Negative residuals tend
to identify stars with strong H$\alpha$ absorption. The rectangular hole is
one of the gaps left by the tiling pattern.

IPHAS provides both images and a source catalogue, as described in
\cite{drew05}. The catalogue contains stars and small resolved objects but not
large scale nebulosities. Coordinates are derived by globally matching the
stellar sources with all-sky astrometric catalogues. The internal precision is
within 0.1 arcsec, and agreement with external catalogues is 0.25 arcsec (e.g.
USNO) to 0.1 arcsec (2MASS). We will use the IPHAS source catalogue below in
the discussion on the proper motion of the central star.

Structure first noted by \cite{tweedy96} to the north-west of the nebula is
now revealed by these observations to be a faint circular closure of the
bright nebula with connecting tails stretching behind. Fig. \ref{iphas-deep}
has scaling chosen to reveal this structure.  From the front of the
filamentary structure to the connecting tails the nebula is 15 arcmin long.
The surface brightness of the tail is less than 1 per cent of that of the
bright filaments.  The tail seems to be stretching away from the nebula in
opposition to an inferred direction of motion which could explain the
one-sided brightening as an interaction with the ISM.

\begin{figure*}
\begin{center}
\includegraphics[width=14cm]{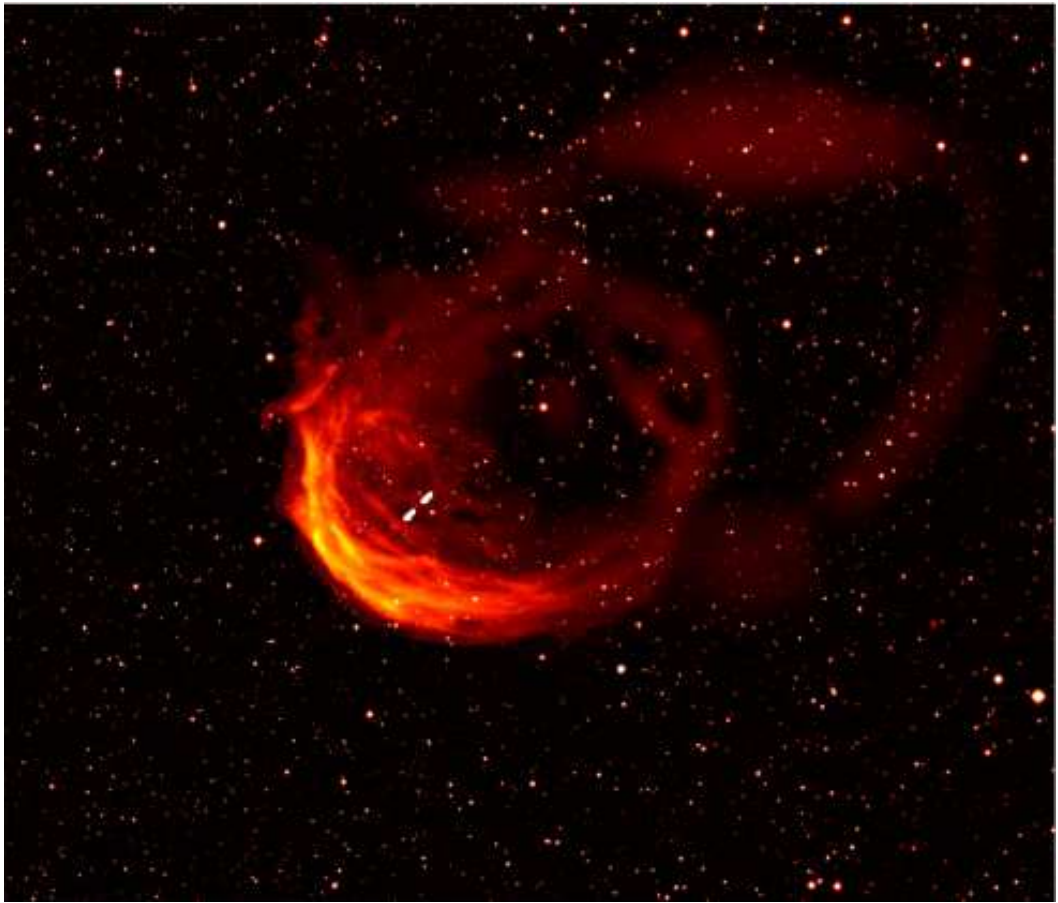}
\caption{An artificial combination of Fig. \ref{iphas-bow} and Fig. \ref{iphas-deep} showing the detail in the bright arc and the faint structure (with some blurring to allow combination of the images). The different IPHAS bands are colour-coded (see text). The H$\alpha$ image includes the continuum. The faint central star is indicated on the image between the markers.}
\label{iphas-vpp}
\end{center}
\end{figure*}

Fig. \ref{iphas-vpp} shows an artificial combination of two three-colour images of the field. One is scaled to show detail in the bright filamentary structure and the other to show the faint structure. The r$\prime$ and the i$^\prime$ bands are coded as blue and green respectively, while H$\alpha$ is coded as both red and yellow. The central star is indicated and is clearly displaced towards the bright arc from the geometric centre of the nebula.

\section{The triple-wind model}

A hydrodynamical code has been used to simulate the interaction with the ISM. The numerical scheme used to solve the hydrodynamics equations uses the second-order Godunov scheme due to \cite{falle91}. The scheme is in 3D cartesian coordinates, fully parallel and includes the effect of radiative cooling \citep{wareing05}.

The numerical domain consists of a cubic grid with 200 cells in each direction and uniform cell spacing. The central mass-losing star is placed at cell coordinates ($50,100,100$) and the simulation is performed in the frame of reference of the star. Mass loss is effected by means of setting the values of density, momentum and energy density in the cells in a spherical region of radius 5$\frac{3}{4}$ cells centred on the star. The radius of the source volume has been chosen by an experimental process whereby it is a balance between producing the most spherical nebula when modelling the stationary ISW model with spherically symmetric winds (i.e. reducing the pixelation of the cartesian grid) and not being large enough to interfere with the results of the simulation. The conditions within this region are reset at the beginning of every timestep to keep driving the wind. Movement through the ISM is parallel to the $x$-axis and material flows in at the ($x=1$) boundary with a positive $x$-velocity. All other boundaries have conditions allowing material to flow out of the domain freely.

The hydrodynamic variables in the source volume are set according to the ISW model of the PN--ISM interaction. This follows the evolution of the AGB and post-AGB phases from the beginning of the AGB phase through the post-AGB or PN phase. In the triple-wind model, this evolution is combined with a constant movement through the ISM.

Mass loss via a stellar wind has been modelled as a spherically symmetric
constant mass-loss rate $\dot{M}$ with constant velocity $v$. In the ISM wind
the density and velocity are constant.

Typical estimates from the literature are used for the wind parameter values.
The post-AGB fast wind parameters are: $\dot{M}_{\rm fw} = 5 \times 10^{-8}$ \mdot, $v_{\rm fw} = 1000$ \kms\ \& $T_{\rm fw} = 5 \times 10^4$ K. The AGB
slow wind parameters are: $\dot{M}_{\rm sw} = 10^{-6}$ \mdot, $v_{\rm sw} =
15$ \kms\ \& $T_{\rm sw} = 10^4$ K. The switch between the AGB wind and the
post-AGB wind is instantaneous and occurs after $10^5$ years of AGB evolution,
which is suffiently long to allow a stable structure to develop. In view of
the still considerable uncertainties on the detailed properties and evolution
of these winds, more detailed temporal variations were not modelled. In
reality, one may expect the AGB wind to show increasing mass-loss rates with
time, whilst the post-AGB may increase in velocity over time.  However, the
current assumptions are found to be sufficient to reproduce the basic
structure of the nebula.  The ISM itself is assumed to be homogeneous with
characteristics of a warm neutral medium, the main constituent of the observed
ISM: $T_{\rm ISM} = 2500$ K \& $n_{\rm H} = 1$ cm$^{-3}$
\citep{burton88}. The gas pressure in all three winds is calculated assuming
an ideal gas equation of state.

\section{Results}

Various speeds of movement have been considered from 0 \kms\ to 175 \kms\ in
steps of 25 \kms. The results of all these models can be found in
\cite{wareing05}. We present here only the case of 125 \kms\ which we find
best fits the general morphology of Sh\,2-188. For comparison with
observations we have considered the locations of the brightened arc and its
faint completing ring, the location of the candidate central star (33 per cent of the
geometric radius upstream of the ring centre), and the location of the
connecting tails downstream of the nebula.

\subsection{Post-AGB structure and evolution}

\begin{figure*}
\begin{center}
\includegraphics[angle=0,width=7cm]{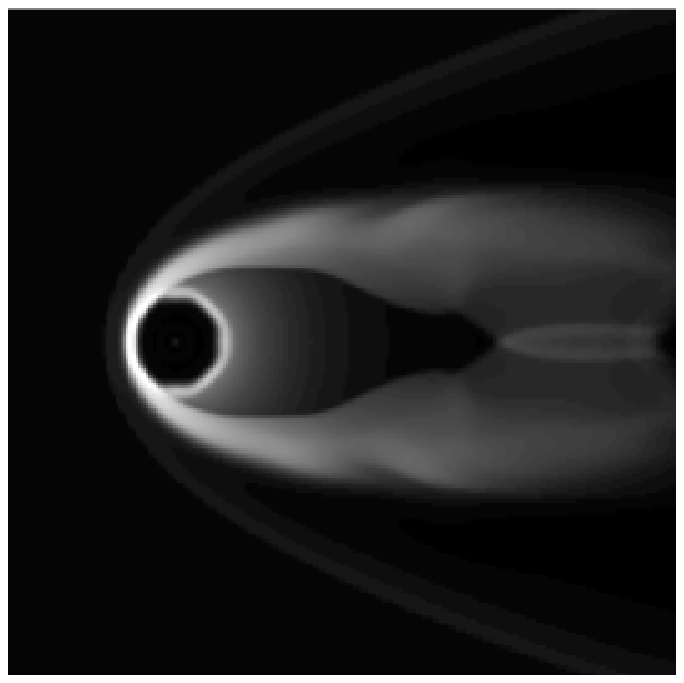} 
\includegraphics[angle=0,width=7cm]{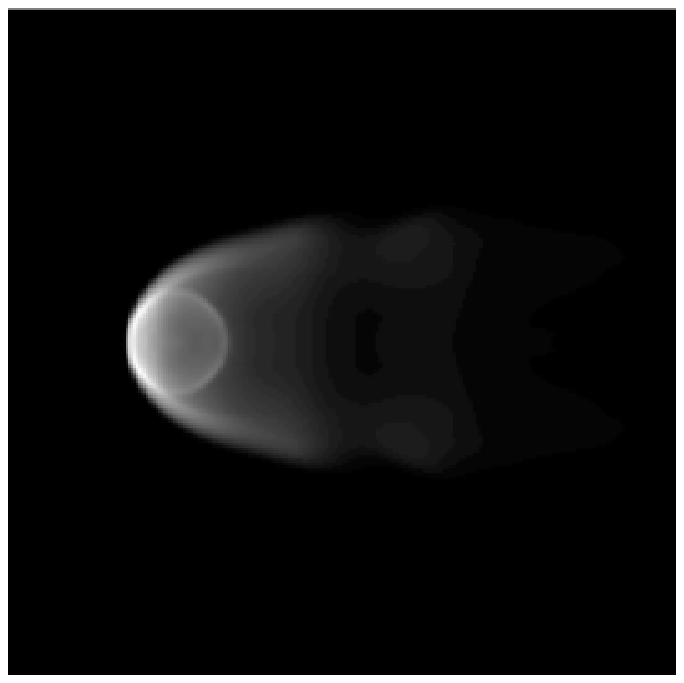} \\
\includegraphics[angle=270,width=7cm]{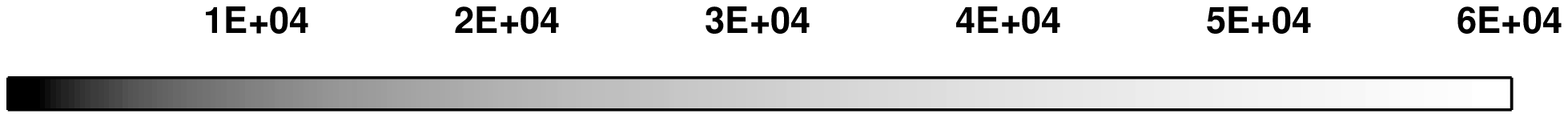} 
\caption{Shown on the left is a slice through the datacube showing the logarithm of density at a point 1,000 years into the PN phase. On the right is a representation of what the nebula might look like in emission. The images are 1 pc on a side. The scaling bar shown refers to the density slice and is presented in \density.}
\label{v125-10000}
\end{center}
\end{figure*}

The left panel of Fig. \ref{v125-10000} shows density on a slice through the
computational domain at ($y = 100$) 1,000 years into the post-AGB (PN) phase.
During the PN phase the wind decreases in density and increases in velocity
resulting in an adiabatic shock that forms the bright shell of the planetary
nebula. Before this point, the nebula would have been inside the AGB wind
bubble created by the AGB wind bow shock. At the depicted time, the nebula has
expanded enough to be interacting with the AGB wind bow shock. Whilst the
nebular shell has a constant temperature all the way round, the upstream part
of the shell is now compressing the AGB wind bow shock and thus has a far
higher density. Velocities in this higher density region range up to 20 \kms. 
The right panel of Fig. \ref{v125-10000} is a synthetic
image obtained by squaring the density and collapsing the datacube along the
line of sight. This is a simple approximation and does not account for
temperature or optical depth. The nebula appears to be brightened upstream
where the higher densities are located. The downstream portion of the nebula
ring is faint yet clear behind the bright arc with the tails connecting
together further downstream.

\begin{figure*}
\begin{center}
\includegraphics[angle=0,width=7cm]{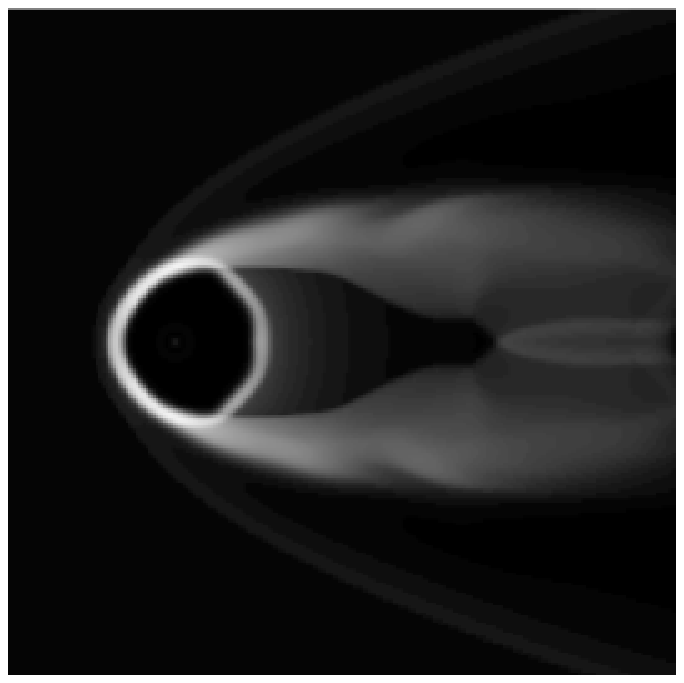} 
\includegraphics[angle=0,width=7cm]{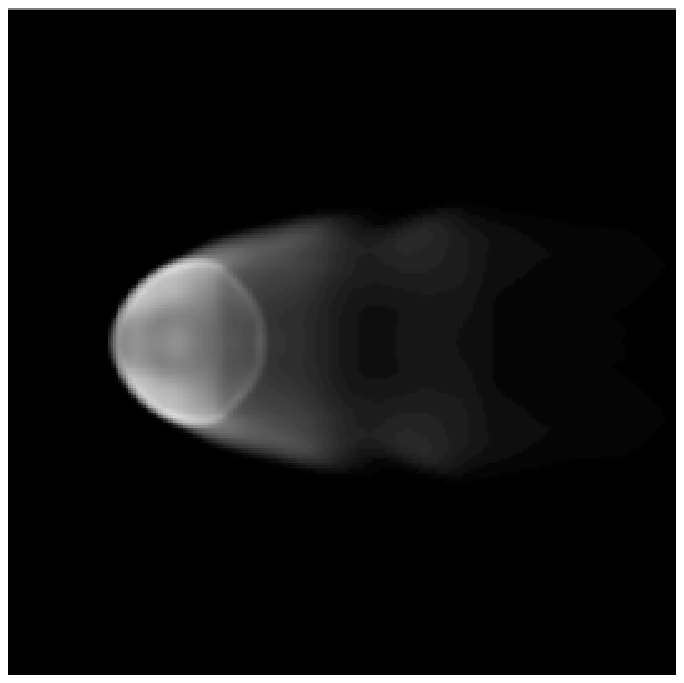} \\
\includegraphics[angle=270,width=7cm]{postAGBdensity-cbar.eps} 
\caption{Shown on the left is a slice through the datacube showing the logarithm of density at a point 2,000 years into the PN phase. On the right is a representation of what the nebula might look like in emission. The images are 1 pc on a side. The scaling bar shown refers to the density slice and is presented in \density.}
\label{v125-20000}
\end{center}
\end{figure*}

Fig. \ref{v125-20000} shows the state of the PN--ISM interaction 2,000 years
into the PN phase. At this stage, the object is most reminiscent of Sh\,2-188.
The upstream arc is brightened, there is a faint completion of the nebular
ring and the tails connect together downstream. Velocities in the upstream bright 
arc range from 20 to 70 \kms. Importantly, the central star
is now displaced 10 per cent of the geometric radius upstream of the geometric
centre. The IPHAS observations of Sh\,2-188 reveal the central star to be
approximately 33 per cent of the geometric radius upstream of the geometric centre.
The faint connection of the ring downstream is the part of the nebula
responsible for this displacement. As the PN shell expands, it is slowed
upstream by interaction with the AGB wind shock whereas downstream, the PN
shell is progressing quicker through undisturbed AGB material, thus causing
the geometric centre to move downstream.

\begin{figure*}
\begin{center}
\includegraphics[angle=0,width=7cm]{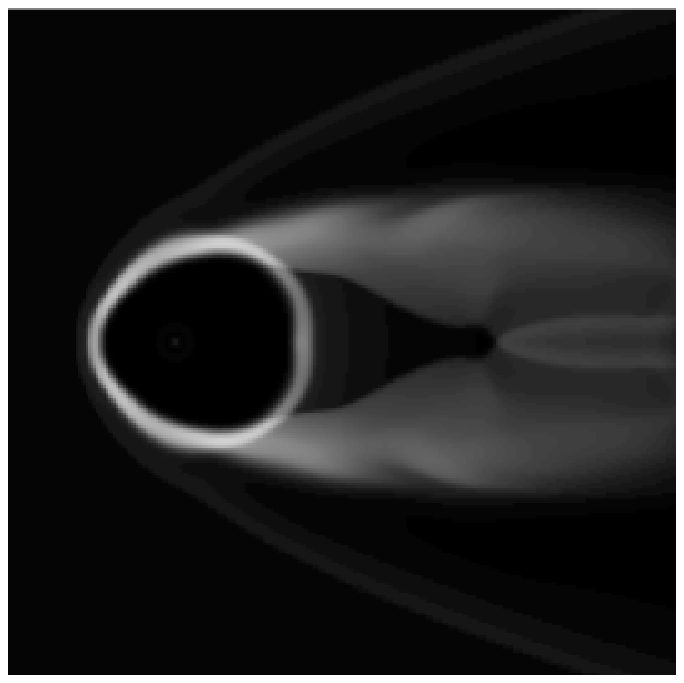} 
\includegraphics[angle=0,width=7cm]{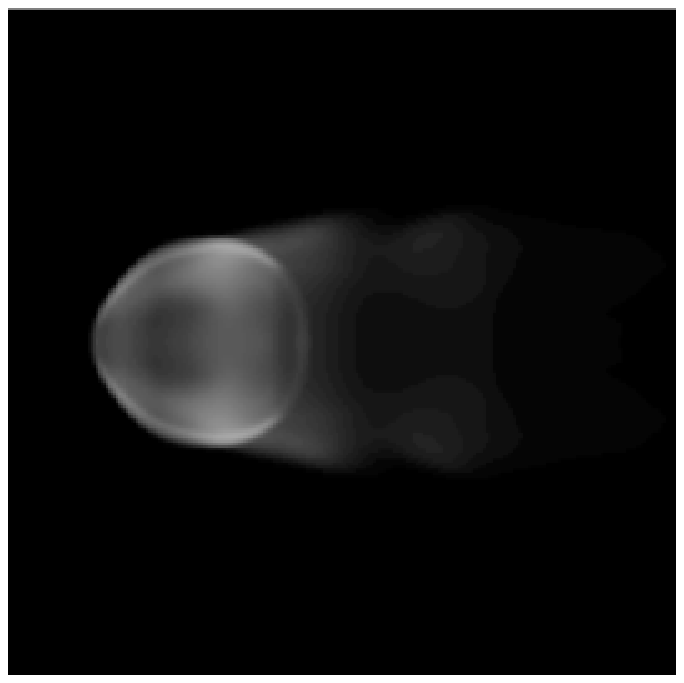} \\
\includegraphics[angle=270,width=7cm]{postAGBdensity-cbar.eps} 
\caption{Shown on the left is a slice through the datacube showing the logarithm of density at a point 3,000 years into the PN phase. On the right is a representation of what the nebula might look like in emission. The images are 1 pc on a side. The scaling bar shown refers to the density slice and is presented in \density.}
\label{v125-30000}
\end{center}
\end{figure*}

The structure of the PN--ISM interaction after 3,000 years of the PN phase is
shown in Fig. \ref{v125-30000}. The nebula is now departing from circular
symmetry and hence appearing less like Sh\,2-188. The brightest regions are
moving downstream to where the PN shell is still interacting with the
shocked AGB material in the tails. The velocities in these regions range between
20 and 50 \kms\ as do the velocities in the shocked high density material at 
the head of the bow shock. The apparent displacement of the central
star is approaching 30 per cent of the geometric radius, similar to the star in
the case of Sh\,2-188. Eventually, after the UV flux of the central star has
turned off and the star has entered the white dwarf cooling regime, the fast
wind will cease.  At this time the whole nebula is swept downstream, and 
the star  begins to desert its nebula as it continues to move at high speed, 
leaving its tail behind.

\subsection{AGB structure and evolution}

\begin{figure*}
\begin{center}
\includegraphics[width=7cm]{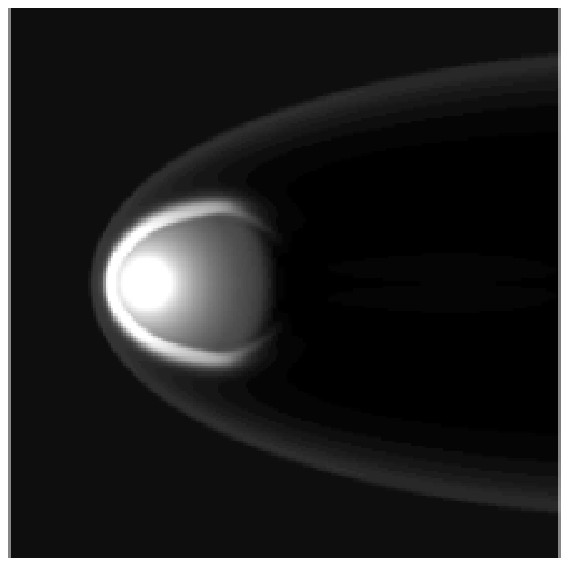} 
\includegraphics[width=7cm]{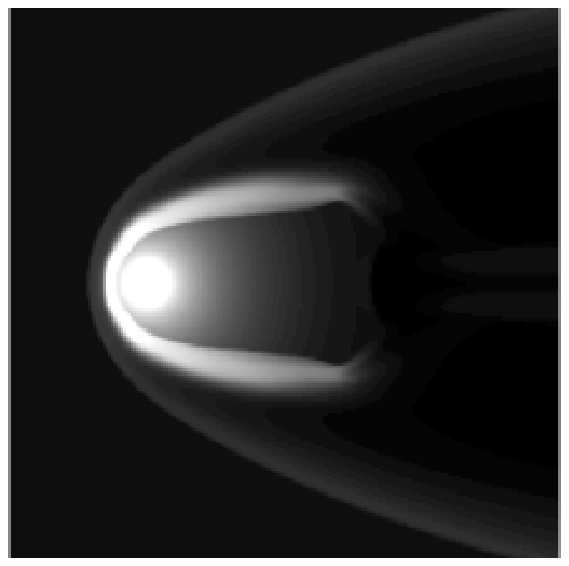} \\
\includegraphics[width=7cm]{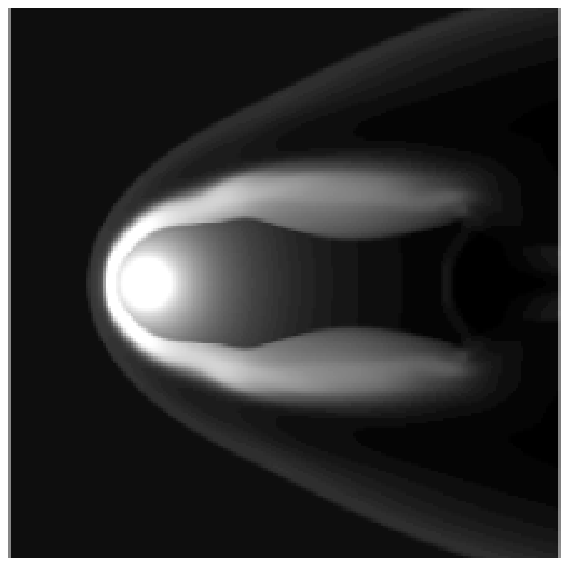} 
\includegraphics[width=7cm]{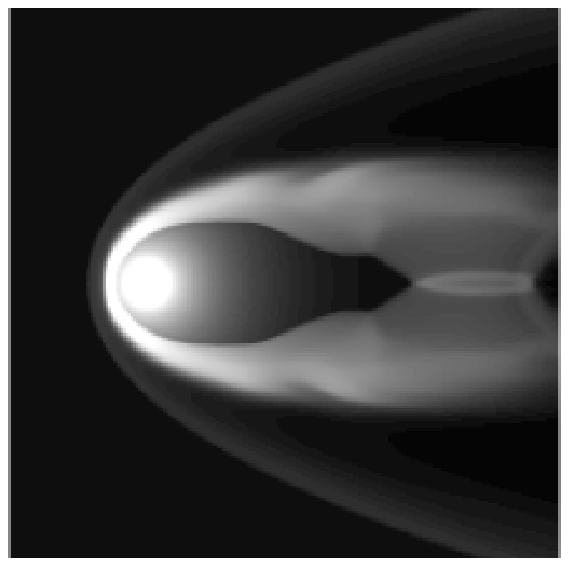} \\
\includegraphics[width=7cm]{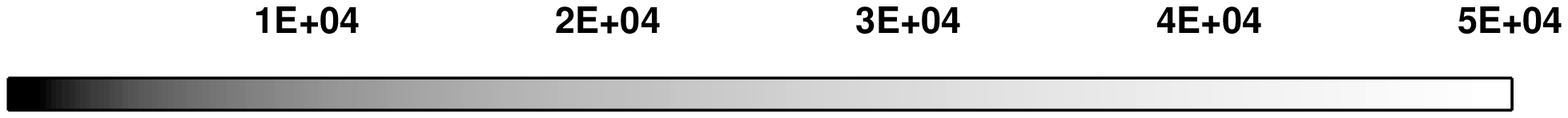}
\caption{Logarithmic density plots showing the development of the bow shock structure and tails during the AGB phase of evolution for a central star with a peculiar velocity of 125 \kms. The images are 1pc on a side. Panel A (upper left) is after 12,500 years, Panel B (upper right) after 25,000 years, Panel C (lower left) after 37,500 years and Panel D (lower right) after 50,000 years. Density is presented in \density.}
\label{AGB}
\end{center}
\end{figure*}

The formation and evolution of the bow shock which so strongly influences the PN shaping is of interest. Fig. \ref{AGB} shows a slice of the computational domain through the position of the central star at various times in the AGB phase. The panels illustrate the development of the bow shock and tail structure. Note that the true structure can be thought of as a rotation of the slices about the symmetry axis through the central star. Compression and deformation of the slow wind against the ISM rapidly forms a bow shock ahead of the star as shown in Panel A, 12,500 years after the start of the AGB phase. In Panel B, 25,000 years into the AGB phase, material stripped by ram pressure from the head of the bow shock is flowing downstream and beginning to form tails behind the nebula.  Eventually, 37,500 years into the AGB phase, the flow collimates these tails downstream of the nebula and they appear to join. At this point in time the structure has reached a stable state with the position of
the reverse shock ahead of the central star being predicted by a simple
ram-pressure-balance argument. The temperature of the shocked ISM material at
the head of the bow shock is approximately 200,000 K. This is in agreement
with strong shock theory which predicts the temperature to be
$ ({3}/{16})\,({m\,v_{\rm ISM}^2}/{k})$, in this case a temperature of
205,500 K. Radiative cooling in the simulation has compressed this region.
The shocked AGB material is ten times more dense and at a temperature of a few
thousand K. The bow shock material cools as it is ram-pressure-stripped
downstream. The structure shown in the fourth panel of Fig. \ref{AGB} is the
state after 50,000 years of the AGB phase, with material clearly flowing back down the tails.
This structure does not change for the next 50,000 years. The injected stellar
wind balances the material moving downstream.

An instability at the head of the bow shock causes material to move down the
tails in waves. This instability can be attributed to vortex shedding at the
head of the bow shock, but the high flow velocity does not allow the vortices
to form fully downstream. In our simulations of lower velocity central stars,
this vortex shedding has been seen to destroy the smooth tail structure. This
effect is not in the scope of this paper and will be discussed at length
elsewhere.

\subsection{Proper motion of the central star}

\begin{table*} 
\caption[A summary of observations of the central star of Sh\,2-188]{A summary
  of observations of the central star of Sh\,2-188 used in the estimation of
  proper motion. The $\Delta$ values give the measurement uncertainties.
 The IPHAS star has a magnitude of $r^\prime = 17.4$ and $r^\prime - {\rm H}\alpha = r^\prime - i^\prime = 0.0$.}
\label{coords}
\begin{center}
\begin{tabular}{|c|c|c|c|c|c|} \hline
epoch & $\alpha$ & $\delta$ & $\Delta \alpha$ & $\Delta \delta$ & Source  \\
& hh\ mm\ ss.s & dd\ $^\prime$\ $^{\prime\prime}$ & $^{\prime\prime}$ & $^{\prime\prime}$ &  \\
\hline
1976.9 & 01\ 30\ 33.113 & 58\ 24\ 50.680 & 0.128 & 0.094 & USNO-B1\ \citep{monet03} \\
1992.8 & 01\ 30\ 33.185 & 58\ 24\ 49.540 & 1.0 & 1.0 & \cite{tweedy96} \\
2003.8 & 01\ 30\ 33.203 & 58\ 24\ 50.290 & 0.165 & 0.165 & IPHAS $r$-band \citep{drew05}\\
\hline
\end{tabular}
\end{center}
\end{table*}

Our simulation reproduces the observed features as described above, if the
central star has a peculiar velocity of 125 \kms. If the star is within 1 kpc
as predicted by various authors (see Table \ref{distances}) and moving as we 
predict at 125 \kms, it should have an
appreciable angular shift on the sky over a relatively short epoch. The
heliocentric radial velocity is reported as $v_r=-26\pm10$ \kms.
Corrections for the Solar motion and Galactic rotation approximately cancel.
Thus, a tangential velocity $v_t=122$ \kms\ with respect to the Local
Standard of Rest (LSR) would be expected.

With this in mind, we have performed a proper motion analysis of the central
star via astrometric methods. The observations used for
this are detailed in Table \ref{coords}.  Initially we looked for a shift
between the IPHAS observations discussed earlier and the observations of
\cite{tweedy96} taken in late 1992 at the 36 inch  Burrell Schmidt
telescope on Kitt Peak, using unbinned IPHAS images with a pixel size of
$\frac{1}{3}$ arcsec. Tweedy and Kwitter's observations have a pixel size of
$2.05$ arcsec, which was insufficient to detect a shift in the position of
the candidate central star.

Following this we found that the central star was listed in the USNO-B1
catalog \citep{monet03} (star number 1484-0053964).  This catalog obtained
data from scans of 7,435 Schmidt plates taken for various sky surveys over the
past 50 years. The candidate central star had been observed in 1976.9 to a
high positional accuracy. We fitted the positions in the catalog to stars
around the nebula up to a radius of $10$ arcmin. The rms of the fit was $0.04$
arcsec. We detected a significant shift of the candidate central star of
$0.8$ arcsec. This shift equates to a proper motion of the central star of
Sh\,2-188 of $26\pm11$ mas\,yr$^{-1}$ in right ascension and of $-15\pm10$ mas\,yr$^{-1}$ in declination. The 1992.8 epoch position encompasses
the expected position of the central star at that time.

This proper motion corresponds to a total displacement in the plane of the sky of $30 \pm 10$ mas\,yr$^{-1}$, with a position angle of $120^{\circ}\pm20^{\circ}$ East through North. The proper motion corrections due to Galactic rotation and due to the Solar motion approximately cancel in this direction.  Assuming the central star is moving at $122\pm25$ \kms, the distance to the nebula is estimated to be $850^{+500}_{-420}$ pc. This distance range is in good agreement with the estimates of \cite{napiwotzki99,napiwotzki01} and \cite{saurer95}. The measured proper motion is in the direction expected from our simulation supporting the identification of the central star of Sh\,2-188. 

Although the agreement of our simulation with observation is encouraging,
this should not be overstated. We have investigated a large parameter range in
velocity, but the range of possible density contrasts is very large and we do
not claim the final result is unique. In fact, now the distance has been
determined, we can also compare the physical size of the nebula with the
simulation: this reveals that the initial solution predicts a far smaller
nebula than is observed. A solution to this discrepancy is discussed in the
next section.

\section{The size of the nebula}

If, as the proper motion combined with the simulation predicts, the central
star is at 850 pc, there is a discrepancy over the physical size of the
nebula between the observations and simulation. The observations show the
nebula is 2.5 pc in diameter whereas in the simulation it is only around one
tenth of this size. The large difference requires a change to the assumed
values of the physical parameters. Since the proper-motion distance and other
distance determinations agree, we keep the velocities unchanged and adjust the
densities.  This approach is also suggested by the fact that observed AGB (super) wind
velocities vary by not much more than a factor of 2, whilst the mass-loss
rates can vary by a factor of 100. A similar argument can be made for the ISM
density versus the peculiar velocity.

A simple rescaling of the simulation allows one to change the densities and
radii, without changing the dimensionless model results.  We wish to
generalise the model holding the velocities in the simulation constant and
increase the size of the nebula by a factor of ten. The size of the nebula is
defined by the size of the AGB wind bubble which is in turn described by a
simple ram pressure balance argument. To move the physical position of the
ram-pressure balance point (reverse shock of the AGB wind bow shock) 10 times
further upstream requires a factor of 100 increase from the ratio of slow-wind
mass loss to ISM density, when keeping velocities constant. Increasing mass,
length and time by a factor of ten decreases density by a factor of 100. Thus
the simulation domain and nebula are ten times larger. The time scales also increase by a
factor of 10: now 500,000 years on the AGB is required for a stable morphology
to develop. This is in line with the 495,000 year AGB evolution predicted
analytically by \cite{vassiliadis93,vassiliadis94}. The surrounding ISM has
dropped in density from 1 cm$^{-3}$ to 0.01 cm$^{-3}$, a value consistent
with the warm ionised medium (WIM) rather than warm neutral medium. 
The rescaling shows that we do not in any way predict the mass-loss rate in
the slow wind, only the ratio between the mass-loss rate and ISM density. The
snapshots of the nebula and AGB evolution presented in the results section
become 10 times older and Sh\,2-188 now appears most like the simulation
after 20 to 30,000 years of PN evolution. Thus, the distance, diameter and age
are all in accordance with \cite{napiwotzki99,napiwotzki01} and
\cite{saurer95}.

Cooling times and cooling effects in the simulation are related to density and so decreasing the density by a factor of 100 alters the cooling. We have rerun the simulation with this rescaling and found that the altered cooling produces the same morphology as the initial simulation, thus the rescaling seems entirely reasonable.

\section{Discussion}


The triple-wind model can reproduce the appearance of the nebula Sh\,2-188.
Accepting the model, the nebula has a peculiar motion with respect to the
local ISM of $125$ \kms. The peculiar motion is mostly in the plane of the sky
with the radial velocity around -25 \kms. The star is located close to the
Galactic plane suggesting the star belongs to the disk population.  The
average transverse motion of a PN-forming central star in the Galactic disk is
40-60 \kms \citep{binney98,dgani98,skuljan99}. The tail of the distribution
extends up to 130 \kms\ \citep{skuljan99, feast00}. The velocity of the
central star is therefore high, yet not unfeasibly high. It would be less
extreme if the star were a thick disk star which have local average velocities
of $50 - 75$ \kms\ \citep{binney98}.

High peculiar velocities are more likely in older stellar populations. This is in conflict to the type-I classification of the nebula: the nitrogen enrichment indicates a more massive progenitor. However, the high [N {\sc ii}]/H$\alpha$ ratio may be caused by shock excitation, and does not necessarily imply an N overabundance. Further evidence of shock excitation is provided by the observation that the sulphur lines are stronger than \hb\ lines at all points where they have been measured \citep{kwitter79}.

The central star of Sh\,2-188 belongs to the tail of the velocity distribution of (older) disk stars. Its rarity is borne out in the unique filamentary, one-sided appearance of the nebula, which is without parallel among known Galactic PNe. The simulations presented in \cite{wareing05} show that a velocity of less than $75$ \kms\ does not result in connecting tails behind the nebula and between $50$ and $100$ \kms\ vortex shedding from the head of the bow shock seems to destroy the smooth tail structure observed in the nebula. A speed in the region of $125$ \kms\ (using increments of 25 \kms) is thus the lowest speed from the simulations at which the morphology of Sh\,2-188 is reproduced.


The observed fragmentation of the bow shock can be understood in terms of the effect predicted by \cite{dgani94,dgani98} (hereafter referred to as DS). They postulated that Rayleigh-Taylor instabilities would fragment a bow shock in the direction of motion. Magnetic fields would suppress certain modes of fragmentation and accentuate others, changing the appearance. Sh\,2-188 is located in the Galactic plane and thus DS interpreted its fragmentation as being an effect of fluid and magnetic field instabilities. Importantly, DS conclude that fragmentation of a substantial part of the halo occurs only for a proper motion of greater than $100$ \kms. This provides a lower limit to the proper motion of Sh\,2-188 and supports the proper motion inferred from the simulation of $125$ \kms. There is no fragmentation in the simulation which could be attributable to a lack of resolution, the lack of magnetic field or a lack of a gravitational field. Further, there is in reality temporal evolution of the slow wind, with thermal pulses at the end of the phase. It is possible that these pulses of high mass-loss could stratify and fragment the bow shock structure before the PN phase. As long as the lifetime of these stratifications allows them to survive late into the PN phase (which is questionable), the fast wind could reveal and further fragment them in a similar manner to the observed filamentary structure.


Spectroscopic data of the bright filaments taken by \cite{rosado82} revealed velocities between 20 and 70 \kms. The velocities in the simulation show the same range in the high density regions at the head of the bow shock which in reality is thought to have fragmented forming the observed filamentary structure. This agreement between the simulation and observations is very encouraging, although no more velocity data on this object are available to us for a more detailed comparison at this time. It should be noted that in our model these velocities are not related to the nebular expansion, but instead to the flow of shocked material from the head of the bow shock back down the tails.


The faint emission completing the bright arc is particularly important. In the simulation this structure has a transitory nature and moves quickly away from the central star: it traces the interaction between the fast post-AGB wind and the slow AGB wind (elsewhere in the nebula this front has merged with the ISM shock). One reason for the fast movement of this shock is the high ratio of post-AGB/AGB momentum which we use.

The distance downstream of the ring from the central star compared to the distance upstream of the bright arc defines the shift of the geometric centre and provides a calibrator for the simulation presented here. The rescaling of the simulation predicts the age of the nebula between 20 and 25,000 years. \cite{napiwotzki99,napiwotzki01} suggests that the star is 22,000 years into the post-AGB phase.

The simulation is not entirely in agreement with the observations. The progressive displacement of a star from the geometric centre of a nebula is an indicator of evolution and in the case of large nebulae, advanced evolution \citep{tweedy96}.  The central star of Sh\,2-188 appears to be approximately 33 per cent of the geometric radius towards the bright arc from the geometric centre. The rescaled simulation indicates that after 20,000 years of PN evolution, the star should appear 10 per cent of the geometric radius upstream of the geometric centre. 10,000 years later, the simulation shows the central star approaching 30 per cent of the geometric radius upstream, in better agreement with the observations. However, by this time, the simulation predicts the bright regions of emission will have begun to shift downstream to where the PN shell is interacting with the tail. A better representation of the time-dependent stellar winds may be needed to resolve this issue. \cite{napiwotzki99,napiwotzki01} locate the star near the start of the white dwarf cooling track. At this point, the wind is expected to cease.  The forward ram shock will no longer be supported by internal pressure and will retreat. The nebula will be blown downstream and disintegrate, and the white dwarf  will appear to move out of its nebula. This phase is not included in the simulation. However, it implies that the Sh\,2-188 morphology has a limited life expectancy.


Interaction with the ISM considerably alters the amount of mass within the circumstellar envelope during the AGB and post-AGB phases. The simulation reaches a point of stability during the life time of the AGB evolution \citep{vassiliadis93,vassiliadis94}: a longer AGB wind does not increase the mass of the shell.  After 20,000 years of the PN phase, an investigation of the (rescaled) simulation reveals that 0.6 \msun\ has been introduced into the spherical region of the nebula. This is made up of 0.5 \msun\ lost by the star and 0.1 \msun\ of ISM material entering the cross section of the nebula.  An integration of the datacube over a roughly spherical region centred on the central star with the nebular radius estimates 0.236 \msun\ remaining within the PN boundary. Thus, 0.364 \msun\ of material has been swept downstream into the tail of the nebula. In fact, the observed tail of the nebula is far older than the nebula itself as it is made up of AGB and ISM material swept downstream. It is likely that most of the ISM material introduced into the stellar region has been swept downstream, but still a minimum of 0.264 \msun\ of stellar material has been mixed with the ISM downstream providing an efficient method of returning stellar material directly to the ISM.

Observations have revealed a missing mass problem in PNe whereby only a small fraction of the mass ejected during the AGB phase is inferred to be present during the post-AGB phase. Stellar evolution calculations predict that stars with initial masses in the range of 1-5 \msun\ will end as PN nuclei with masses around 0.6 \msun. Most of the mass is lost during the AGB phase and should be easily observable as ionised mass during the PN stage. However, observations of Galactic PNe have revealed on average only 0.2 \msun\ of ionised gas. Several PNe have been shown to have embedded molecular clumps which could contain much of this missing mass and these clumps have shown to survive ablation by stellar winds and therefore could survive into the PN phase \citep{pittard05}. The stripping of mass downstream of a moving PNe as investigated here also provides an attractive solution to this missing mass problem; the Dumbbell nebula appears to have clumps of material surrounding the bright nebula inside what would appear to be an AGB wind bow shock \citep{meaburn05} combining both of these solutions to the missing mass problem.

Observations \citep{napiwotzki99,napiwotzki01} have indicated that the central star of Sh\,2-188 has a mass of 0.6 \msun. These observations, in combination with the mass injected into our simulation and the addition of mass lost prior to the superwind phase (a typical estimate for low-mass stars being $\lsim 0.5$ \msun) allows us to estimate a mass of $\sim 1.5$ \msun\ for the progenitor star.


\cite{villaver02} discussed evolution of AGB stars and the interaction of
their dust shells with the ISM. Typically, AGB stars form large shells (up to
4 pc in diameter) around them. Evidence for such an AGB--ISM interaction front
was found by \cite{zijlstra02}, who found a M3 III AGB star surrounded by such
a detached shell of 4 pc diameter. The ISM influence is defined by the
density and hence pressure in the ISM, which drops off exponentially from $n_{\rm H} = 2$ cm$^{-3}$ \citep{spitzer78} at a scale height of 100 pc \citep{binney98}.
The regular shells can be influenced strongly by the movement through the
surrounding ISM.  The AGB evolution of the central star of Sh\,2-188 shows a
structure extensively altered from these regular dust shells. Much of the AGB
material has been swept downstream.

At a distance of 850 pc, the nebula is 60 pc below the Galactic plane.  We
would expect the ISM density in the region of Sh\,2-188 to be around $n_{\rm H} = 2$
cm$^{-3}$ so it would seem our rescaling choice of $n_{\rm H} = 0.01$ cm$^{-3}$ is at odds
with this. However, large regions of the Galactic plane in between spiral arms
are dominated by such low-density gas, including the volume around the Sun
\citep{frisch95}. In fact, the NE2001 model \citep{cordes02}, used commonly 
in the Pulsar community to predict the galactic distribution of free electrons, 
predicts an electron density in the direction of Sh\,2-188 of $n_{\rm e} = 0.013\ 
{\rm cm}^{-3}$ at a distance of 850 pc. This is in very good agreement with our 
prediction, but it should be stressed that this is only one 
choice of AGB mass-loss rate to ISM density ratio. A higher mass-loss rate would
increase the ISM density, although this would have implications for the mass in
the nebular region.  The direction of motion indicates the star is moving
almost entirely in Galactic longitude and has travelled 64 pc during the AGB
and post-AGB phases.  Therefore the assumption of constant ISM properties
during the AGB and post-AGB evolution of the central star may not be correct.

\section{Conclusions}


We have successfully reproduced the morphology and the available kinematic data of Sh\,2-188 and understood its formation in terms of the `triple wind' model. Following the AGB evolution of the central star has been crucial in fitting the whole structure, in particular the tail behind the nebula which is comprised of purely AGB material. The triple-wind model of Sh\,2-188 has predicted a velocity of the candidate central star of 125 \kms. Velocities from spectroscopic data on the bright filaments \citep{rosado82} are in agreement with velocities measured from the simulation. A proper motion study of the central star has shown it to be moving at $30.0 \pm 10.0$ mas\,yr$^{-1}$ in the direction of the head of the bright arc. The combination of these two measurements has resulted in estimates of $D = 850^{+500}_{-420}$ pc, $d \sim 2.5$ pc and $t_{\rm PN} = 22,500 \pm 2,500$ years. These estimates are in agreement with the distance and age estimates of \cite{saurer95} and \cite{napiwotzki99,napiwotzki01}. The prediction of ISM density in the vicinty of Sh\,2-188 is also in agreement with the NE2001 model of \cite{cordes02}. The triple wind model explains the geometric displacement of the central star and indicates that the faint closure of the bright arc is a transitory structure which evolves downstream of the nebula. The PN--ISM interaction has caused $\approx \frac{2}{3}$ of the mass expected in the region of the star to be swept downstream providing a solution to the missing mass phenomenon in PN and a valuable way of mixing ISM and stellar material several pc downstream of the central star.

The success of the triple-wind model to fit this extreme object gives confidence in our ability to fit objects with lower speeds. It is now clear that the outer halo structure of PNe contain the effects of ISM interaction and should not be modelled as stand-alone structures. Further, the ISM interaction is an important method of mixing stellar material back into the ISM. The next generation of telescopes, particularly ALMA, will be able to reveal cool dust structure in the universe and shed light on circumstellar AGB material. Further simulations considering temporal evolution of the stellar winds, magnetic fields and/or gravity may shed light on why we have not reproduced the observed fragmentation in Sh\,2-188.

\section*{Acknowledgments}

CJW was supported by a PPARC research studentship. The numerical computations
were carried out using the Jodrell Bank Observatory COBRA supercomputer.  The
paper uses data obtained from the IPHAS survey, carried out at the Isaac
Newton Telescope (INT) on the island of La Palma. The INT is operated by the
Isaac Newton Group in the Spanish Observatorio del Roque de los Muchachos of
the Instituto de Astrofisica de Canarias. The authors acknowledge the IPHAS 
collaboration, consisting of, in addition to authors of this paper: A. Aungwerowijt, M. Barlow,
R. Corradi, D. Frew, B. G\"{a}nsicke, P. Groot, A. Hales, E. Hopewell, M. Irwin, 
C. Knigge, P. Leisy, D. Lennon, A. Mampaso, M. Masheder, M. Matsuura,
L. Morales-Rueda, R. Morris, Q. Parker, S. Phillips, P. Rodriguez-Gil,
G. Roelofs, I. Skillen, D. Steeghs, Y. Unruh, K. Viironen, J. Vink,
N. Walton, A. Whitham and A. Zurita. Useful comments from Sir Francis Graham-Smith, Ian Browne and Alan Pedlar at Jodrell Bank led to a useful discussion with Michael Kramer, also at Jodrell Bank, over electron densities. Finally, thanks to the referee, John Dyson, for excellent constructive comments which led to the supportive comparisons with available spectroscopic data.

\label{lastpage}

\end{document}